# van der Waals Torque in 2D Materials Induced by Interaction between Many-Body Charge Density Fluctuations

Zepu Kou,[1] Yuquan Zhou,[1] Zonghuiyi Jiang,[1] Alexandre Tkatchenko,[2,*] and Xiaofei Liu[1,†]

[1]State Key Laboratory of Mechanics and Control for Aerospace Structures, Key Laboratory for Intelligent Nano Materials and Devices of the Ministry of Education, Nanjing University of Aeronautics and Astronautics, Nanjing, 210016, P. R. China

[2]Department of Physics and Materials Science, University of Luxembourg, L-1511 Luxembourg City, Luxembourg

E-mail: [*]alexandre.tkatchenko@uni.lu; [†]liuxiaofei@nuaa.edu.cn

**Van der Waals torque determines the relative rotational motion between anisotropic objects, being of relevance to low-dimensional systems. Here we demonstrate a substantial torque between anisotropic two-dimensional materials that arises from the interaction between many-body charge density fluctuations, exceeding by twenty-fold the torque computed with atom-pairwise models. The dependence of torque on the disorientation angle, $\theta$, in the form of $-\sin(2\theta)$, the positive correlation between torque and in-planar dielectric anisotropy, the linear relation between torque and area, and the decaying torque with increasing separation are rediscovered using the fully atomistic many-body dispersion model. Unlike continuum Casimir-Lifshitz theory, the advantage of the molecular theory relies on describing the collective torque and the effects of atomic details on an equal footing. These findings open an avenue for incorporating quantum fluctuation-induced torque into molecular modeling, being instrumental to the design of nanoelectromechanical systems and the understanding of rotational dynamics of anisotropic layered materials.**

Casimir and van der Waals interaction stemming from quantum and thermal fluctuations is relevant to a wide range of disciplines, including performance of micro-/nanoelectromechanical systems [1,2], surface tension of liquids [3,4], adhesive force between biological molecules [5], and mechanical properties of condensed matters [6-9]. For macro- or mesoscopic objects, the Casimir energy is derived from boundary condition-dependent zero-point energy of electromagnetic fields, as done within the Casimir-Lifshitz theory [10,11]. In nanoscale systems, the Casimir force instead corresponds to the intermolecular van der Waals (vdW) force that can be described by empirical force fields or charge density-based methods [12-17]. As predicted by the Casimir-Lifshitz theory, parallel plates with in-planar dielectric anisotropy experience not only an interlayer force but also an interlayer torque [18,19]. The dielectric anisotropy-induced torques between a liquid crystal and birefringent materials and that between two rutile $TiO_2$ (001) surfaces have been detected recently [20,21].

Ever since the seminal prediction of Casimir/vdW torque by Parsegian and Weiss [18], the Casimir-Lifshitz theory has been the main tool to describe the torque-related phenomena. Modulations of Casimir/vdW torque via changing dielectric functions of interacting materials and of liquid environment have been extensively explored [22,23]. An alternative strategy to induce Casimir/vdW torque via utilizing anisotropic geometric patterns has been theoretically proposed [24,25]. A peculiar retardation effect that enhances the torque and a distance-dependent sign reversal of torque were revealed [26,27]. General Casimir/vdW torques between dielectrically isotropic objects without spherical symmetry were proposed by using a numerical Casimir-Lifshitz theory or the proximity force approximation [28,29]. The great success of the continuum theory relies on its ability to capture all the collective polarization effects, including the synergistic contribution by electronic and ionic degrees of freedom, the contribution by magnetic response, the effect of finite light speed, and particularly the dielectric anisotropy [30,31]. On the other hand, because of the assumption of continuum dielectric media, atomic details important to nanoscale systems are completely neglected by the Casimir-Lifshitz theory. Hence, the continuum theory is restricted to extended dielectric or metallic objects at relatively large separations, precluding investigations of realistic phenomena in a wide range of nanoscale systems where an atomistic treatment is required as we will show below.

Recent years witness the boosting of vdW methods for atomic systems, including empirical molecular force fields [32,33], density functional theory (DFT)-based models [12-16], and even machine learning-based methods [34-36]. It is of paramount importance to obtain consistencies between the molecular theories and the continuum theory. This has been achieved for inter-object force: A summation of pairwise (PW) interatomic forces can lead to correct scaling laws of vdW force between parallel semi-infinite plates, and between parallel two-dimensional (2D) layers [37,38], in agreement with the continuum theory. However, molecular simulation has seldom been applied to explore the dielectric anisotropy-induced torque in nanoscale systems, let alone to derive the torque scaling laws.

In this Letter, we demonstrate that the vdW torque between anisotropic 2D materials arises from the interaction between many-body charge density fluctuations. The interlayer torque between black phosphorus (BP) monolayers computed by the many-body dispersion (MBD) model [39] exhibits an angular dependence in the form of $-\sin(2\theta)$, with $\theta$ being the disorientation angle. The positive correlation between torque and in-planar dielectric anisotropy, the linear relation between torque and area, and the decaying torque with increasing separation are rediscovered via the MBD, being in excellent consistency with the continuum theory. In contrast, neglecting the many-body coupling, neither the PW theory with or without three-body term nor the two-point vdW density functional could describe the interlayer torque. By revealing vdW torque synergistically contributed by contact area and dielectric anisotropy, vdW torque jointly induced by geometric and material anisotropies, and effect of surface adsorbate, we highlight that the advantage of the molecular theory against the continuum theory relies on its ability to describe the collective torque and the effects of atomic details on an equal footing. These findings shed a light on incorporating quantum fluctuation induced-torque into molecular modeling, giving access to a multitude of vdW torque-related phenomena at nanoscale systems.

Noncovalent dispersion interaction is important to 2D materials that have no surface dangling bonds, influencing the interlayer cohesion, the interlayer electronic and phononic couplings, the elastic constant along the normal direction, and the interaction with substrate or liquid/gas environment. For this reason, extensive theoretical and experimental efforts have been devoted to exploring the interlayer force between 2D materials [40-42]. Recent upsurge of exploiting Moiré pattern-related properties further underlines the urgency of understanding the angular dependence of interlayer interaction [43,44], especially for anisotropic 2D materials [45-48].

Since the Casimir torque between BP monolayers has been studied by Thiyam *et al.* using the Casimir-Lifshitz theory [27], AB-stacked circular disks of BP monolayers are chosen as the testing platform for the performance of commonly used vdW methods in describing interlayer torque. As displayed in Fig. 1a and Fig. S1 of the Supplementary Material [49], the upright projection of the top BP disk coincides with the bottom BP disk. The disorientation angle-dependent interlayer energy, $\Delta E(\theta)$, is calculated by rotating the top disk around the normal axis passing through its center. The benefit of such treatment is to rule out the influence of contact area, defined as the area of upright projection of the top disk onto the bottom one. The interlayer torque is then calculated from the energy, as $M = -d\Delta E/d\theta$.

Figure 1b presents the torque as a function of disorientation angle. For BP disks with a diameter ($D$) of 60 Å and separated by 20 Å, the PW theory results in a maximum torque density of ~20 nN·m·m$^{-2}$, which is lower than the MBD torque as discussed below by more than twenty-fold. Even worse, the maximum PW torque appears on $\theta$ of ~$(2n+1)\pi/12$, with $n$ being an integer, in conflict with the fact that the maximum torque between birefringent plates appears on $\theta$ of $(2n+1)\pi/4$ (see Fig. S2 of the Supplementary Material for the detailed PW results [49]). As illustrated in Fig. 1b and Fig. S3 of the Supplementary Material [49], including the Axilrod–Teller–Muto three-body term via the D3 method [50] does not amend the deficiencies. Figure S4 of the Supplementary Material [49] shows that the vdW-DF2 method [51] is unable to describe the interlayer torque either. These tests confirm that molecular simulation of vdW torque is a challenging task.

The maximum MBD torque density in the above case is ~500 nN·m·m$^{-2}$, being in the same order of magnitude to the Casimir-Lifshitz result estimated from Thiyam *et al.*'s data [27] (see Table SI in the Supplementary Material for the detailed comparison [49]). According to Somers *et al*'s experimental data of Casimir torque between bulk materials and the $d^{-2}$ scaling law [20], the maximum torque density on the liquid crystal (5CB) exerted by LiNbO$_3$ (YVO$_4$) at a separation of 20 Å is estimated to be 420 (1077) nN·m·m$^{-2}$. The lower torque between BP monolayers relative to that between bulk 5CB and YVO$_4$ can be rationalized, since BP is atomically thin and has an in-planar dielectric anisotropy weaker than those of 5CB and YVO$_4$ (see Table SII in the Supplementary Material for materials polarizability anisotropy [49]). As illustrated in Fig. 2a, the BP disks have a maximum interlayer energy on $\theta$ of $\pi/2$ and $3\pi/2$, namely, the system is most unstable when the zigzag direction of the top disk is perpendicular to that of the bottom one. This $\Delta E$-$\theta$ relation leads to a maximum MBD torque on $\theta$ of 48º, 133º, 228º and 313º. The slight deviations from the angles of maximum Casimir-Lifshitz torque could be a result of inter-edge interactions between the disks [52]. To confirm that the MBD torque originates from materials in-planar dielectric anisotropy, the MBD torque between two circular graphene disks separated by 20 Å is shown in Fig. 1a and Fig. S5 of the Supplementary Material [49]. Indeed, without in-planar dielectric anisotropy, the maximum energy difference between graphene disks with $D$ of 60 Å is merely 3.47×10$^{-6}$ meV, being five orders of magnitude lower than that between BP disks. Figure S6 of the Supplementary Material [49] shows that changing the stacking pattern into AA style does not induce any evident change of interlayer torque, again verifying that the MBD torque is a dielectric anisotropy-induced collective effect.

To reproduce the linear relation between interlayer torque and area of interacting plates as assumed within the continuum theory, Fig. 2a illustrates the angular dependence of interlayer MBD energy of BP disks with different diameters (60, 80 or 100 Å). The as-obtained area-normalized $M$-$\theta$ relations almost coincide with each other (Fig. 2b), implying that there is no small-size effect of torque density at least for BP disks with $D > 60$ Å and the linear relation between torque and area works for atomically thin materials.

Figure 2c and 2d present the distance dependences of interlayer energy difference and of interlayer torque, respectively, for a BP disk with $D$ of 80 Å interacting with a BP disk with $D$ of 114 Å. The purpose of using a relatively larger bottom disk is to attenuate any possible inter-edge interaction. It is found that both the maximum energy difference ($\Delta E_{max}$) and the maximum torque ($M_{max}$) decrease with separation, $d$. In the range from 10 to 30 Å, the slope of the $\log(M_{max})$-$\log(d)$ relation varies from −2.2 to −2.6 (see Fig. 2e). The scaling law of torque different from that for parallel semi-infinite plates given by the Casimir-Lifshitz theory ($M_{max} \sim d^{-2}$) [18] is mainly due to the atomically thin thickness of BP. Nevertheless, the slope of the $\log(\Delta E_{max})$-$\log(d)$ relation at $d$ of 30 Å (−3.9) is close to the exponent of power law for interlayer energy between two strictly-2D layers (−4) [37], justifying the distance dependence by the MBD.

One striking observation is that the angles of maximum torque in the above case perfectly agree with those predicted by the continuum theory. Especially, the $M$-$\theta$ relation can be well fitted in the form of $-\sin(2\theta)$ (see Fig. 2d), just identical to that firstly proposed by Barash using the Casimir-Lifshitz theory [19]. Such an angular dependence of interlayer torque even persists for the BP disks separated by merely 1 nm.

According to the Casimir-Lifshitz theory, the torque between two homogeneous plates is positively correlated to the square of dielectric anisotropy ratio [19]. By comparing the torque between BP disks and that between SiP$_2$ disks [47], we reveal that an analogous rule works for the MBD torque. Table I illustrates the ratio of in-planar polarizability anisotropy, defined as

$$\gamma = \frac{\varepsilon_{\infty u} - \varepsilon_{\infty v}}{\varepsilon_{\infty v} - 1}, \tag{1}$$

where $\varepsilon_{\infty u}$ and $\varepsilon_{\infty v}$ are the static electronic dielectric functions along two principle axes. $\gamma_{SiP2}$ and $\gamma_{BP}$ derived using the HSE06 exchange-correlation functional [53] are 0.277 and 0.158, respectively, leading to $\gamma_{SiP2}^2/\gamma_{BP}^2$ of 3.07 and $\gamma_{BP}\gamma_{SiP2}/\gamma_{BP}^2$ of 1.75. As expected, the ratios of maximum energy difference and maximum torque of SiP$_2$ disks to those of BP disks ($\Delta E_{max@SiP2}/\Delta E_{max@BP} = 2.47$, $M_{max@SiP2}/M_{max@BP} = 2.32$) are close to $\gamma_{SiP2}^2/\gamma_{BP}^2$. Similarly, the ratios of the interlayer interactions of heterogeneous BP-SiP$_2$ disks to those of BP disks ($\Delta E_{\text{max@BP-SiP2}}/\Delta E_{\text{max@BP}}$ = 1.80, $M_{\text{max@BP-SiP2}}/M_{\text{max@BP}}$ = 1.81) are close to $\gamma_{\text{BP}}\gamma_{\text{SiP2}}/\gamma_{\text{BP}}^2$. See Fig. S7 of the Supplementary Material for the detailed results of SiP$_2$-SiP$_2$ and BP-SiP$_2$ [49].

It must be mentioned that, the polarizability anisotropy ratio of 2D material and the ensuing torque are sensitive to $\beta$, a parameter used in the MBD to modulate the truncation of short-range Coulomb potential. To avoid the double counting of short-range correlation energy already contained within the exchange-correlation functional, the MBD only describes the long-range part [54]. Namely, a Fermi-type damping function is applied, expressed as

$$f(r_{ij}) = 1/(1 + e^{-6[r_{ij}/\beta(R_{\text{vdW}}^i + R_{\text{vdW}}^j) - 1]}), \quad (2)$$

where $r_{ij}$ is the distance between atom $i$ and atom $j$, $R_{\text{vdW}}$ is the vdW radius, and $\beta$ is a rescaling factor. Hence, before calculating the torque, one shall choose a reasonable $\beta$. The key is that if only the self-consistent screening [39] under the same damped Coulomb potential reproduces the DFT ratio of in-planar polarizability anisotropy, the as-applied $\beta$ would result in a correct MBD torque. As listed in Table I, $\beta$ of 0.62 and 0.53 are suitable for BP and SiP$_2$, respectively.

In realistic situations, the interlayer interaction can be affected by a wealth of atomic details, including geometric shape, surface adsorbates, wrinkles, bending deformation, stacking manner, and finite-size effects. Because of the assumption of continuum medium, these details cannot be captured by the Casimir-Lifshitz theory. In contrast, the MBD possesses the advantage of describing the collective torque and the effects of atomic details on an equal footing, which will be highlighted by taking three effects of geometric shape or of surface adsorbate as examples.

Firstly, the vdW torque synergistically contributed by varying contact area and materials anisotropy is captured by the MBD. Figure 3a shows two square BP flakes separated by 20 Å, whose upright projections coincide with each other on $\theta$ of 0º. As the top BP fake rotates around its center, the contact area varies with $\theta$ (see Fig. 3b), thus contributing to the interlayer torque. As evidenced by the PW result, without in-planar dielectric anisotropy, the energy minimum of the square flakes would appear on $\theta$ of 0, $\pi/2$, $\pi$ and $3\pi/2$, corresponding to the angles of maximum contact area. However, the MBD energy on $\theta$ of $\pi/2$ or $3\pi/2$ is lifted by 0.24 meV relative to that on $\theta$ of 0 or $\pi$ (see Fig. 3d), due to the in-planar dielectric anisotropy of BP. The resulting $M$-$\theta$ relation by the MBD is distinct from that by the PW theory. While the maximum PW torques solely due to the contact area effect are identical on different angles, the largest MBD torque on $\theta$ of 160º is higher by 900 nN·m·m$^{-2}$ than the second largest on $\theta$ of 70º. Such effect is due to the additive effect of contact area-related and anisotropy-induced torques on $\theta$ of 160º and inversely the cancellation effect on $\theta$ of 70º. Notably, although strategies of utilizing contact area-dependent energy have been applied to provide restoring forces for sheared graphite [55], sheared double-wall nanotube [56], and a finite-size plate suspended above a finite-size substrate [57], the synergistic vdW torque due to contact area and materials anisotropy has never been demonstrated.

Secondly, we present a new type of vdW torque that originates jointly from geometric and material anisotropies. As mentioned in the introduction, interlayer torque can be induced by either material or geometric anisotropies, but the joint effect remains unknown. As shown in Fig. S8 of the Supplementary Material [49], the interlayer MBD energy between an armchair graphene ribbon (40 Å × 15 Å) and a BP disk ($D = 80$ Å) exhibits an angular dependence, resulting in a maximum torque density of 2082.37 nN·m·m$^{-2}$ at a separation of 20 Å. Since the in-planar dielectric response of graphene is isotropic, the observed torque can be considered as a joint effect of geometric anisotropy of graphene ribbon and material anisotropy of BP. We note that a similar effect shall exist in continuum systems, but whose discovery would require a further development of fluctuating surface current approach dedicated for anisotropic material of arbitrary shape [28].

Thirdly, the MBD can be applied to understand how a surface adsorbate influences the interlayer interaction. Upon the adsorption of a graphene flake consisting of seven carbon hexagons on one of the circular BP disks (Fig. 3c), the interlayer MBD torque between the BP disks separated by 20 Å is almost unchanged (Fig. 3f). However, as shown in Fig. 3g, the interlayer force is enhanced upon the appearance of adsorbate.

To provide an understanding on the origin of interlayer torque from the view point of many-body charge fluctuations, the density of states of plasmonic MBD modes in an isolated BP disk and the overall average dipole moments along $x$, $y$ and $z$ axes of each mode are reported in Fig. 4a. Due to the in-planar anisotropy, the dipole moments along the $x$ and $y$ directions are unequal, which leads to the angular dependent interlayer coupling. Figure 4b presents the energy of MBD mode in the coupled disks relative to that in the isolated disks, as a function of mode number. In the full energy range, the interlayer coupling results in downward or upward energy shifting; the shorter the interlayer separation is, the stronger is the energy shifting. As the top disk is disoriented by $\pi/2$, the magnitude and the site of energy shifting are modified, especially for those near energy $E_2$ (labeled in Fig. 4c). Among the three energies with evident shifting upon interlayer coupling ($E_1$, $E_2$ and $E_3$ in Fig. 4d), the polarization vector of $E_2$ mode is most anisotropic, with the armchair component being 2.4 times of the zigzag component.

The incapability of the PW theory can be understood from the angular dependence of average interatomic distance between the top and the bottom circular BP disks (see Fig. S9 of the

Supplementary Material [49]). The variation of average interatomic distance is smaller than $1 \times 10^{-5}$ Å, rationalizing the insensitivity of PW energy to $\theta$. In the case of the two-point vdW-DF, the dielectric function and the polarization operator adopted to calculate correlation energy are derived from electron density without considering the dielectric anisotropy [15,58]. Consequently, it is not promising to describe the interlayer torque via the vdW-DF. In principle, the interlayer torque can be predicted using correlated methods such as the random phase approximation and the quantum Monte Carlo simulation [40,59], though the computational cost would be unaffordable.

To conclude, we have demonstrated that the quantum fluctuation-induced torque between anisotropic two-dimensional materials arises upon considering the many-body coupling of charge density fluctuations. Being able to faithfully reproduce the interlayer torque and related rules and even to predict new torque-related phenomena, the many-body dispersion model provides a fully atomistic alternative to the macroscopic Casimir-Lifshitz theory, and alters the situation that dielectric anisotropy-induced torques have been largely overlooked in previous force field- and charge density-based molecular modeling. Since the interlayer interaction is not only important to fundamental properties of 2D materials and of their one-dimensional derivatives, but also crucial to the construction of heterogenous devices, describing Casimir/vdW torque and effects of atomic details on an equal footing by a molecular theory would benefit the understanding of Moiré structure constructed with anisotropic materials and the design of rotational nanoelectromechanical devices.

X. L. is supported by the National Natural Science Foundation of China (Grants No. 12072152 and No. 11702132) and the Fundamental Research Funds for the Central Universities (Grants No. NS2024003). A. T. is supported by the European Research Council (ERC-AdG Grant FITMOL).

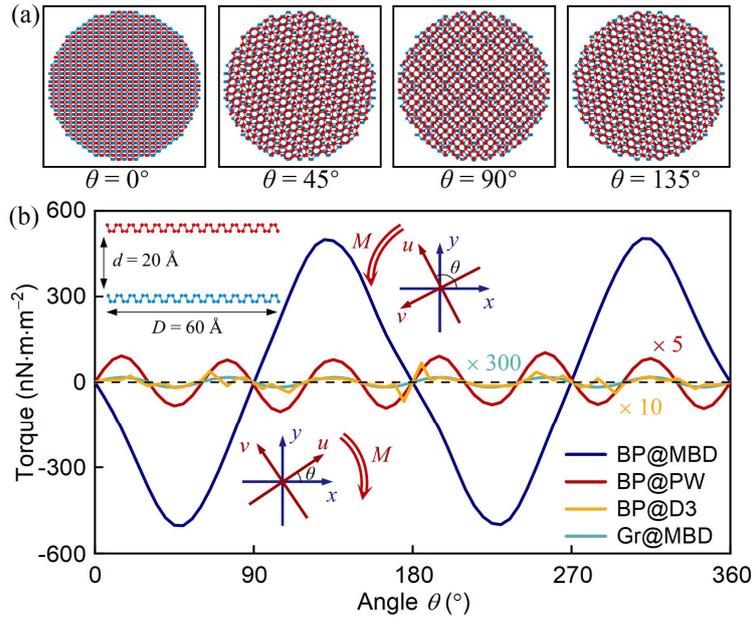

FIG. 1. (a) Illustration of AB-stacked circular black phosphorus (BP) disks used for testing the performance of vdW methods in describing dielectric anisotropy-induced torque. In the four configurations, the bottom disk is fixed with the armchair direction being aligned along the $x$ axis, and the top disk is rotated around its center by corresponding disorientation angles. See Fig. S1 of the Supplementary Material [49] for the detailed atomic structure of circular BP disk. (b) VdW torques per unit area between the BP disks with a diameter of 60 Å and separated by 20 Å, and between two graphene (Gr) disks with the same diameter and separation. The presented PW data of BP, D3 data of BP and MBD data of graphene are amplified by 5, 10 and 300 times, respectively. The insets illustrate the torque ($M$) on the top disk exerted by the bottom disk on two different disorientation angles.

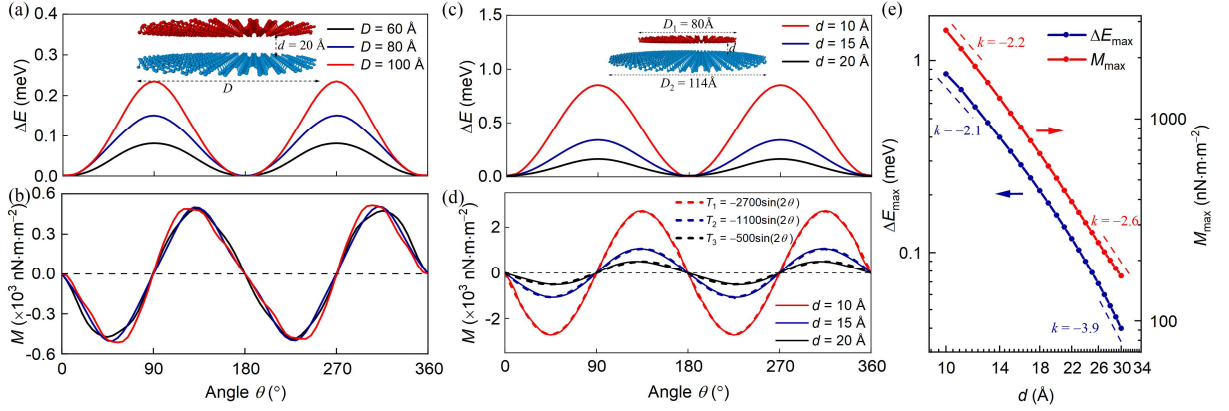

FIG. 2. Dependences of energy and torque on the area and the separation of circular BP disks, computed by the MBD. (a) Relative interlayer energy ($\Delta E$) between BP disks with varying diameters ($D$ = 60, 80 or 100 Å) under a separation of 20 Å, as a function of disorientation angle. The inset illustrates the atomic structure and geometric parameters. (b) Torque per unit area for the three cases shown in (a). (c) Relative interlayer energy of a BP disk with $D$ of 80 Å interacting with a BP disk with $D$ of 114 Å, under varying separation ($d$ = 10, 15 or 20 Å). (d) Torque per unit area for the three cases shown in (c). The area of the top disk is considered as the contact area. The dashed curves are obtained via a fitting in the form of $M \sim -\sin(2\theta)$. (e) log-log plots of the maximum interlayer energy difference and the maximum interlayer torque density as a function of separation.

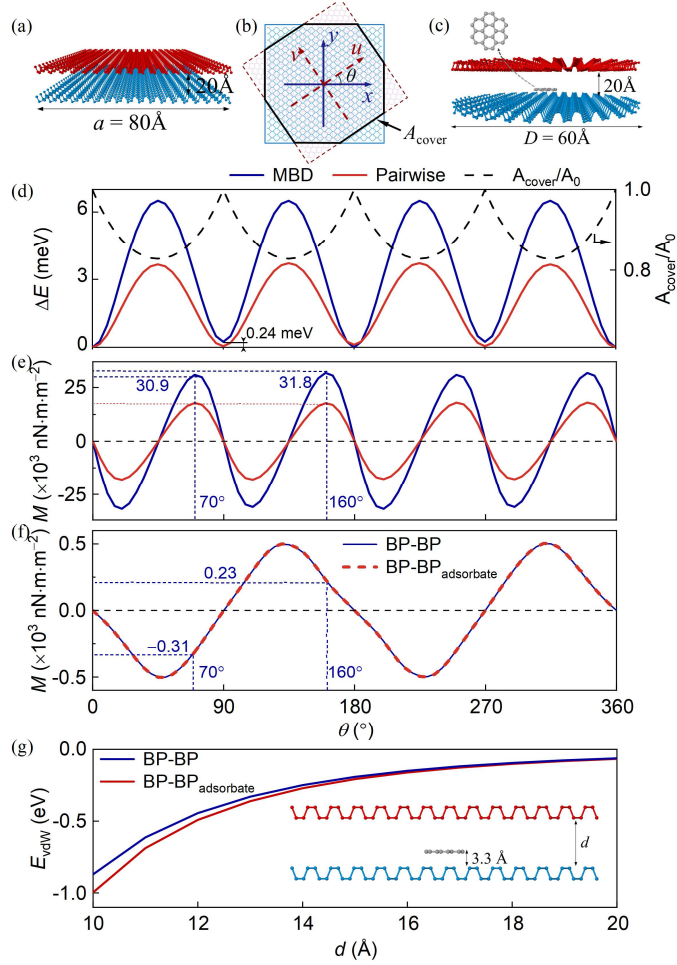

FIG. 3. Effects of atomic details on the interlayer energy and torque between BP monolayers. (a) Illustration of parallel square BP flakes with a side length of 80 Å. (b) Illustration of the angular dependence of contact area ($A_{cover}$). (c) Illustration of a circular BP disk with an adsorbed graphene flake of seven carbon hexagons, interacting with another circular BP disk. See Fig. S1 of the Supplementary Material [49] for the detailed atomic structures of square BP flakes and of circular BP disks with adsorbate. (d) Relative interlayer energy between the square BP flakes, as a function of disorientation angle. The dashed line shows the contact area as a function of angle. (e) Torque per unit area of the square BP flakes. (f) Torque per unit area of the circular BP disks with (dashed line) or without (solid line) the adsorbate. The vertical dashed lines are used to underline the different signs of dielectric anisotropy-induced torque on the angles of maximum torque shown in (e) (70º and 160º). (g) Interlayer energy between the circular BP disks with or without the adsorbate on a fixed angle ($\theta = 0º$), as a function of separation.

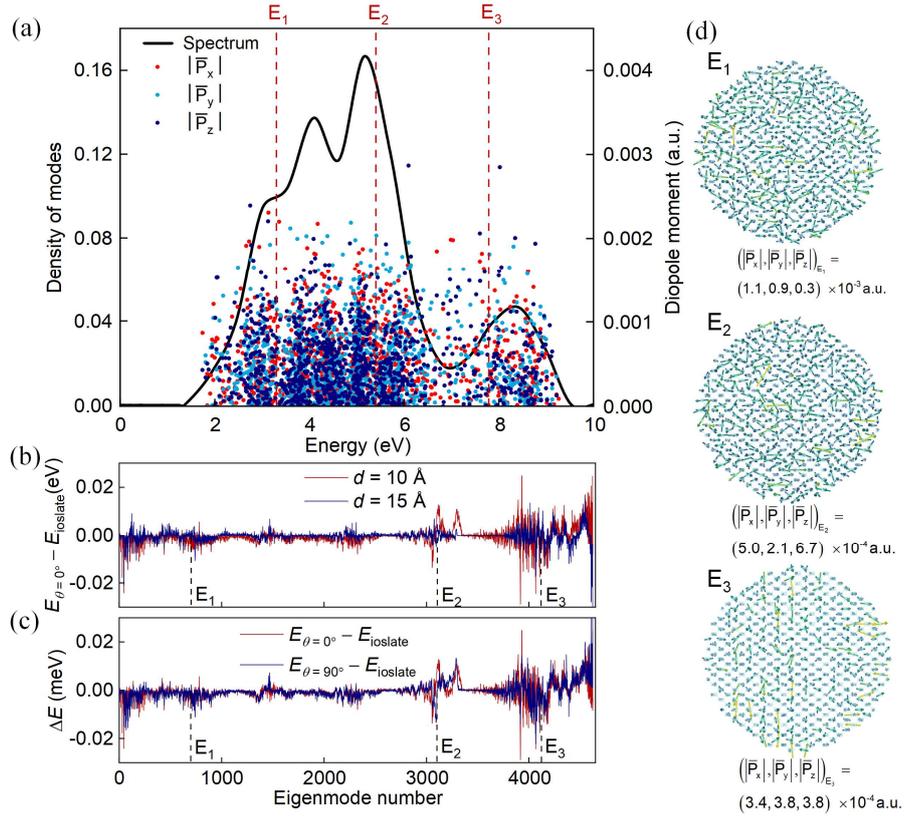

FIG. 4. MBD modes of charge density fluctuations. (a) MBD energy spectrum and overall average dipole moments along *x*, *y* and *z* axes, of a single BP disk with *D* of 60 Å. (b) MBD mode energy of the coupled disks (*d* = 10 or 15 Å, $\theta$ = 0º) relative to that of the isolated disks, as a function of mode number. (c) MBD mode energy of the coupled disks (*d* = 10 Å, $\theta$ = 0º or 90º) relative to that of the isolated disks. (d) Top views of the charge fluctuation modes at energies $E_1$, $E_2$, and $E_3$. The arrows denote the dipole moments of atoms.

Table I. In-planar polarizability anisotropy of 2D materials, maximum interlayer energy difference and maximum interlayer torque density between circular disks with $D$ of 80 Å and separated by 20 Å, and $\beta$ applied in the MBD. The armchair graphene ribbon is 40 Å in length and 15 Å in width. See Fig. S1 of the Supplementary Material [49] for the detailed results of $SiP_2$-$SiP_2$, BP-$SiP_2$, Gr-Gr and Gr ribbon-BP.

| 2D layers | $(\varepsilon_u - \varepsilon_v)/(\varepsilon_v - 1)$ | $\Delta E_{max}$ (meV) | $T_{max}$ (nN·m·m$^{-2}$) | $\beta$ |
|---|---|---|---|---|
| BP-BP | 0.158 | 0.15 | 487.38 | 0.62 |
| $SiP_2$-$SiP_2$ | 0.277 | 0.37 | 1129.77 | 0.53 |
| BP-$SiP_2$ | 0.158/0.277 | 0.27 | 879.76 | 0.62/0.53 |
| Gr-Gr | 0 | $6.82\times10^{-6}$ | - | 0.56 |
| Gr ribbon-BP | 1.985/0.158 | 0.07 | 2082.37 | 0.56/0.62 |

# Supplementary Material for "van der Waals Torque in 2D Materials Induced by Interaction between Many-Body Charge Density Fluctuations"

Zepu Kou,[1] Yuquan Zhou,[1] Zonghuiyi Jiang,[1] Alexandre Tkatchenko,[2,*] and Xiaofei Liu[1,†]

[1]State Key Laboratory of Mechanics and Control for Aerospace Structures, Key Laboratory for Intelligent Nano Materials and Devices of the Ministry of Education, Nanjing University of Aeronautics and Astronautics, Nanjing, 210016, P. R. China

[2]Department of Physics and Materials Science, University of Luxembourg, L-1511 Luxembourg City, Luxembourg

E-mail: *alexandre.tkatchenko@uni.lu; †liuxiaofei@nuaa.edu.cn

## Contents

# 1. Atomic structures used in the calculations

Figures S1a and S1b illustrate the atomic structures of AB-stacked circular BP disks. Both the top and bottom disks were built from a pristine BP monolayer by cutting off atoms outside a circle of given radius. On the top disk, the center of circle resides in the center of a P six-member ring. On the bottom disk, the center of circle resides in the middle of a P-P bond aligned along the armchair direction. The AA-stacked circular BP disks were built in a similar way, excepted that both the centers of top and bottom disks reside in the center of a P six-member ring, as illustrated in Figs. S1c and S1d.

Figure S1e illustrates the atomic structures of AB-stacked square BP flakes. The horizontal and vertical sides are aligned along the armchair and the zigzag directions, respectively. The centers of the top and the bottom squares reside in the center of a P six-member ring and the middle of a P-P bond, respectively. The BP flakes were built by cutting off atoms outside the square.

Figure S1f illustrates the atomic structures of a circular BP with an adsorbed graphene flake consisting of seven carbon hexagonal rings. The center of the graphene flake resides in the center of BP circle.

# 2. Details of the MBD model

The MBD calculations were performed using the pyMBD code [1]. In the MBD model, the long-range correlation energy $E_{vdW}^{LR}$ was calculated via the coupled fluctuating dipole model [2,3]. The Hamiltonian of MBD is defined as

$$H_{MBD} = -\frac{1}{2}\sum_{i=1}^{N}\nabla_{d_i}^2 + \frac{1}{2}\sum_{i=1}^{N}\omega_i^2 d_i^2 + \sum_{i>j}^{N}\omega_i\omega_j\sqrt{\alpha_i^0 \alpha_j^0}\, d_i \mathbf{T}_{ij} d_j, \qquad (S1)$$

where $\alpha_i^0$ and $\omega_i$ are the static dipole polarizability and the characteristic excitation frequency of atom $i$, respectively. $d_i$ represents the mass-weighted displacement of the $i$th oscillator from its equilibrium distance. The dipole-dipole interaction tensor between atoms $i$ and $j$ reads $\mathbf{T}_{ij} = \nabla_{r_i} \otimes \nabla_{r_j} v(|r_i - r_j|)$, where $v$ is a modified Coulomb potential at the interatomic separation $r_{ij} = |r_i - r_j|$. To avoid double counting of the short-range correlation energy already contained within the exchange-correlation functional, the MBD only describes the long-range part. Namely, a Fermi-type damping function is used to truncate the short-range Coulomb potential, expressed as

$$f(r_{ij}) = 1/(1 + e^{-6[r_{ij}/\beta(R_{vdW}^i + R_{vdW}^j) - 1]}), \qquad (S2)$$

where $r_{ij}$ is the separation between atom $i$ and atom $j$, $R_{vdW}$ is the vdW radius of atom, and $\beta$ is a rescaling factor. The MBD energy is then computed as the difference between the zero-point energies of the interacting and the noninteracting systems

$$E_{vdW}^{LR} = \frac{1}{2}\sum_{i=1}^{3N}\bar{\omega}_i - \frac{3}{2}\sum_{j=1}^{N}\omega_j, \tag{S3}$$

where $\bar{\omega}_i$ is the Hamiltonian eigenvalues and $\omega_j$ is eigen frequency of free atom $j$. Since chemical bonding and especially charge transfer between atoms in an ionic compound have influence on the vdW parameters of materials, the $C_6$ coefficients and polarizabilities derived from the nonlocal many-body dispersion method (MBD-NL) [4] (see Table SIII) were applied for the MBD calculation of vdW torque.

The in-planar dielectric anisotropy was evaluated using the density functional theory with the HSE06 exchange-correlation functional [5], as encoded in the FHI-aims code [6]. The calculations were carried out for primitive cells using a $k$-point mesh of 14×10×1 for BP and of 15×5×1 for $SiP_2$. A vacuum layer of 25 Å was used to avoid any spurious interaction between periodic images. The cutoff energy for the plane wave basis was set to 450 eV. To determine a suitable $\beta$ in the coupled fluctuating dipole model that can reproduce the DFT dielectric anisotropy ratio, the self-consistent screening (SCS) model under the same damped Coulomb potential as the MBD is used to evaluate the in-planar dielectric anisotropy ratio. In the SCS, the molecular polarizability tensor is derived from a Dyson-like self-consistent equation

$$\boldsymbol{\alpha}^{SCS}(i\omega) = \boldsymbol{\alpha}^0(i\omega) - \boldsymbol{\alpha}^0 \mathbf{T} \boldsymbol{\alpha}^{SCS}(i\omega), \tag{S4}$$

where $\boldsymbol{\alpha}^0$ is the bare polarizability tensor, $\boldsymbol{\alpha}^{SCS}$ is the self-consistently screened polarizability tensor and $\mathbf{T}$ is the long-range interaction matrix.

Figure S11 shows the damping functions with $\beta$ of 0.62 and 0.83. When $\beta$ is 0.83, the Coulomb interaction between oscillators on two nearest P atoms are almost truncated, which is the reason why the SCS under a relatively large $\beta$ fails to reproduce the in-planar polarizability anisotropy.

## 3. Details of the pairwise theory, the D3 and the vdW-DF calculations

In the pairwise theory, the interlayer vdW energy is calculated as the summation of pairwise interatomic potentials

$$E_{pairwsie} = -\sum C_6 R_{ij}^{-6}, \tag{S5}$$

where $i$ and $j$ indices denote atoms in the top and the bottom monolayers, respectively. For a fair compassion with the MBD, the $C_6$ coefficients and polarizabilities derived from the MBD-NL [4] were used for the calculation of pairwise energy. Note that the damping function as used in the Tkatchenko-Scheffler theory [7] does not affect the interlayer interaction here, since the separation considered is much larger than the sum of two vdW radii.

The D3 energy was calculated using the Dftd3 code [8,9]. The dispersion energy $E_{disp}$ is the sum of the two- and three-body contributions

$$E_{disp} = E^{(2)} + E^{(3)}, \qquad (S5)$$

where $E^{(2)}$ and $E^{(3)}$ are dispersion energies of two-body and three-body contributions, receptively. The two-body term at long range is

$$E_{disp} = -\frac{1}{2} \sum_{A \neq B} \sum_{n=6,8} s_n \frac{C_n^{AB}}{r_{AB}^n}, \qquad (S6)$$

where $C_n^{AB}$ denotes the averaged (isotropic) $n$th-order dispersion coefficient for the atom pair A-B, and $r_{AB}$ is the internuclear distance. $s_n$ is an exchange-correlation functional-dependent scaling factor. The Becke-Johnson damping is used for the dispersion correction.

The vdW-DF2 [10] energy was calculated using the VASP code [11]. Since the DFT calculation for large disks as used in the pyMBD calculations (more than 3000 atoms) is unaffordable, the vdW-DF2 energy of smaller BP disks was calculated. Each BP disk contains ~100 atoms, with a diameter of 17 Å. To avoid the spurious interactions between nearby periodic images, a large supercell (35 Å × 35 Å × 30 Å) was applied.

## 4. Supplementary figures and tables

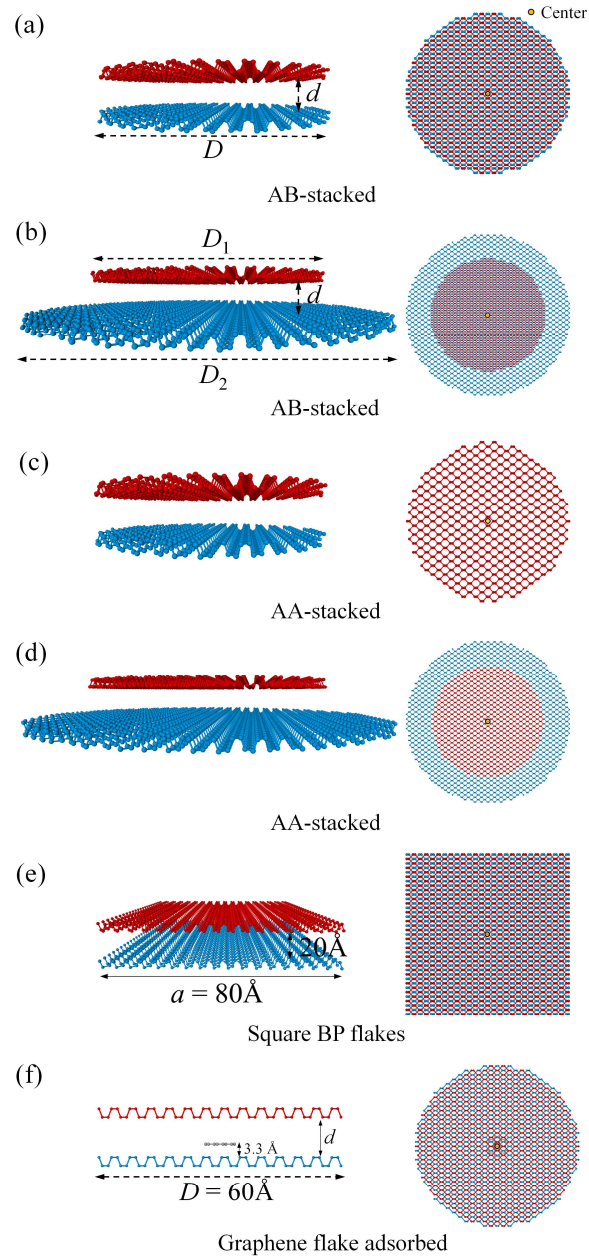

FIG. S1. Atomic structures used in the MBD calculations. (a) Illustration of the atomic structure of AB-stacked circular BP disks with $D$ of 60 Å. The yellow dot in the right panel denotes the center of the BP disks. (b) Illustration of AB-stacked BP disks with different diameters. (c) Illustration of the atomic structure of AA-stacked circular BP disks with $D$ of 60 Å. (d) Illustration of AA-stacked BP disks with different diameters. (e) Illustration of the atomic structure of AB-stacked square BP flakes. (f) Illustration of the atomic structure of AB-stacked circular BP disks with an adsorbed graphene flake of seven carbon hexagons.

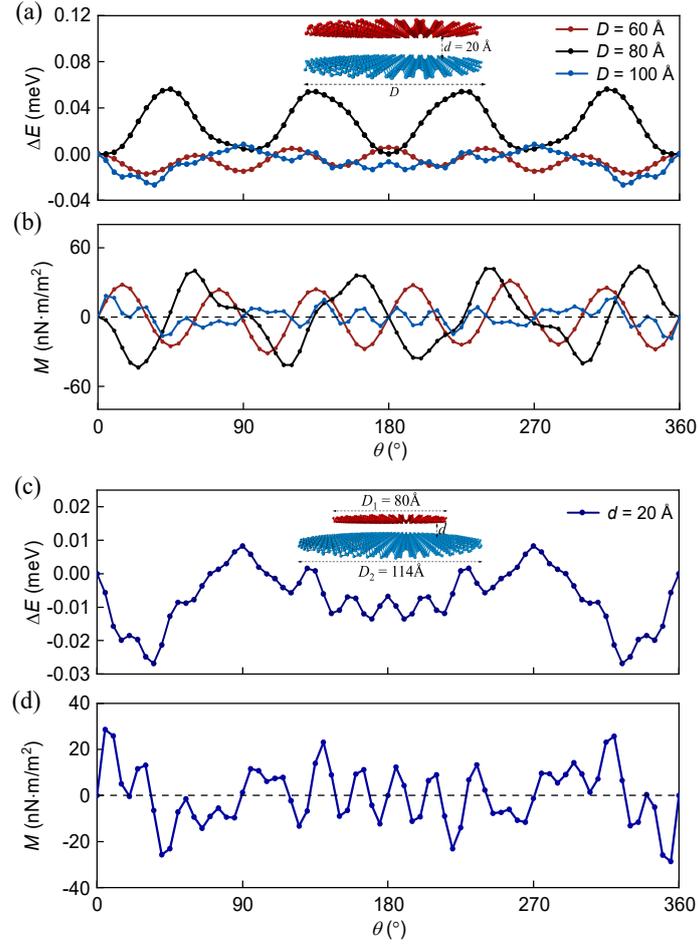

FIG. S2. (a) Pairwise (PW) relative interlayer energy between circular BP disks with $D$ of 60, 80 or 100 Å and separated by 20 Å, as a function of disorientation angle. (b) Torque per unit area derived from the PW energy shown in (a). (c) PW relative interlayer energy of a circular BP disks with $D$ of 80 Å interacting with a BP disk with $D$ of 114 Å, as a function of disorientation angle. (d) Torque per unit area derived from the PW energy shown in (c).

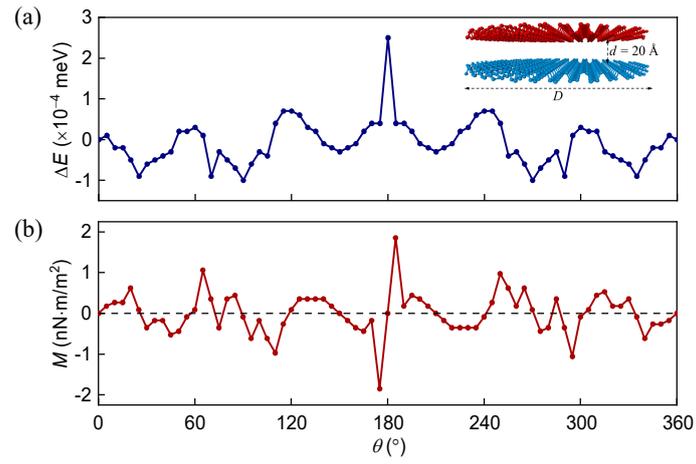

FIG. S3. (a) Relative interlayer energy between circular BP disks with $D$ of 60 Å and separated by 20 Å as a function of disorientation angle, calculated by the D3. (b) Torque per unit area derived from the D3 energy.

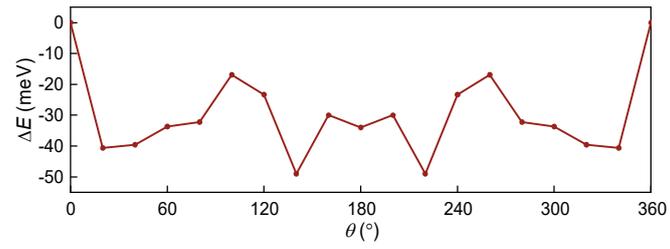

FIG. S4. Relative interlayer energy between BP disks with a diameter of ~17 Å and separated by 10 Å as a function of disorientation angle, calculated by the vdW-DF2 [10].

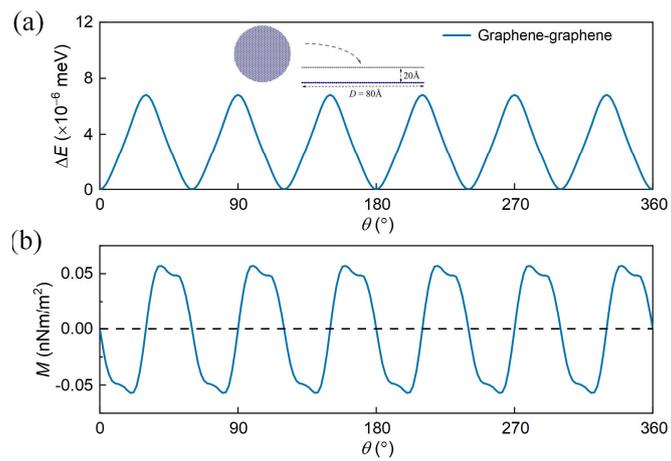

FIG. S5. Relative interlayer energy (a) and vdW torque density (b) between circular graphene disks with $D$ of 80 Å and separated by 20 Å as a function of disorientation angle, calculated by the MBD.

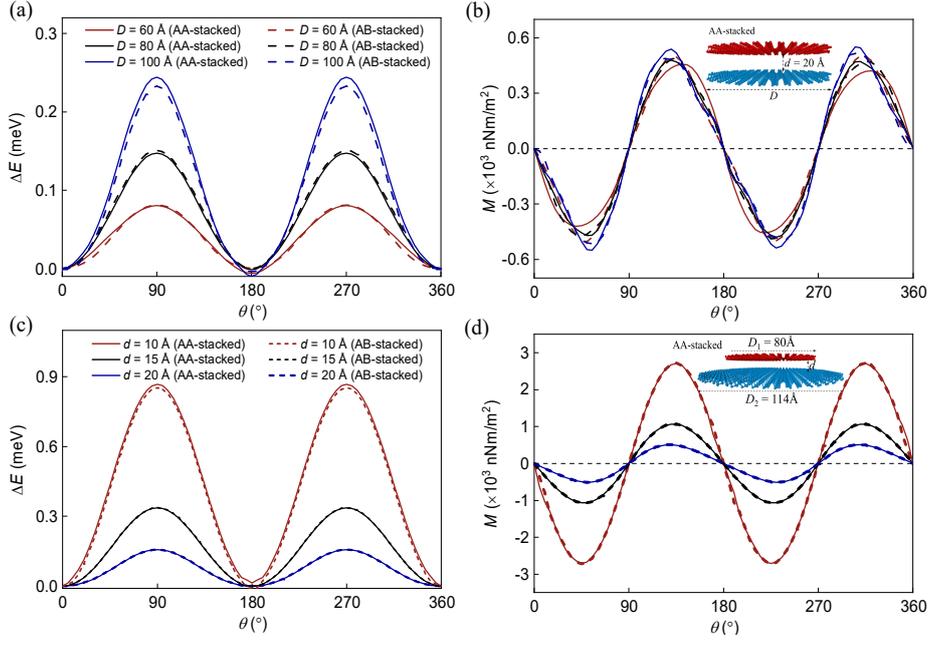

FIG. S6. Relative interlayer energy and vdW torque density between AA-stacked BP disks, in comparison with the results for AB-stacked BP disks. (a) Relative interlayer energy between BP disks with $D$ of 60, 80 or 100 Å, calculated by the MBD. (b) Torque per unit area for the cases shown in (a). (c) Relative interlayer energy of a BP disk with $D$ of 80 Å interacting with a BP disk with $D$ of 114 Å, calculated by the MBD. (d) Torque per unit area for the cases shown in (c).

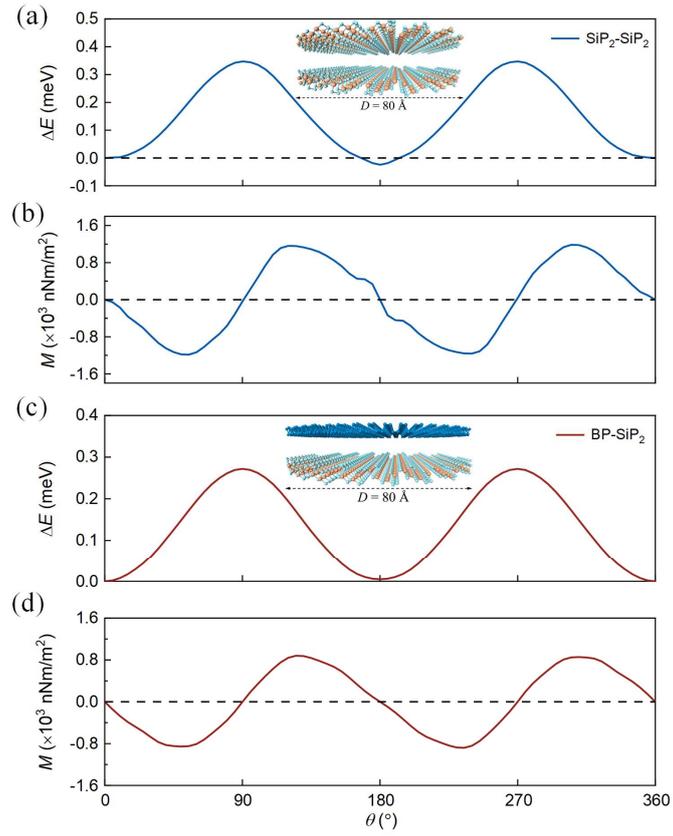

FIG. S7. (a) Relative interlayer energy between circular SiP$_2$ disks with $D$ of 80 Å and separated by 20 Å, calculated by the MBD. (b) Torque density between circular the SiP$_2$ disks. (c,d) The same as (a,b), but for the interlayer interaction between BP and SiP$_2$ disks.

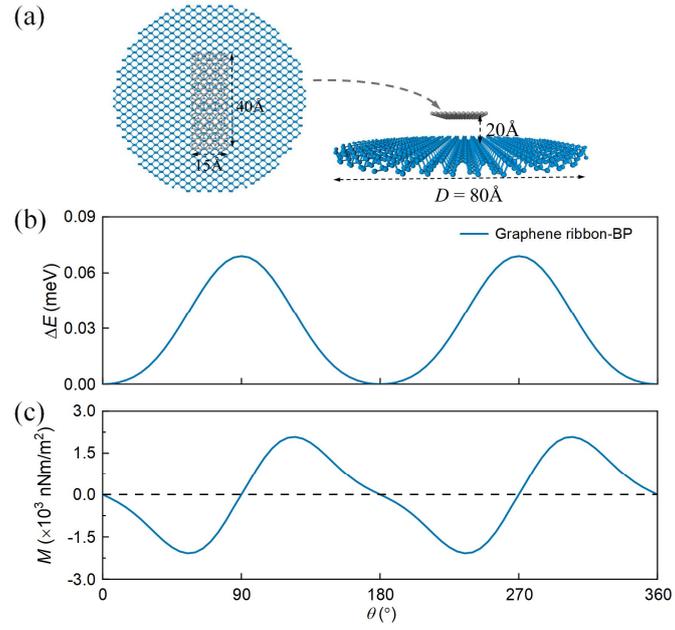

FIG. S8. (a) Illustration of the atomic structure of an armchair graphene ribbon (40 Å × 15 Å) and a circular BP disk with $D$ of 80 Å under a separation of 20 Å. (b) Relative interlayer energy between graphene ribbon and BP disk calculated by the MBD. (c) Torque density between the graphene ribbon and the BP disk.

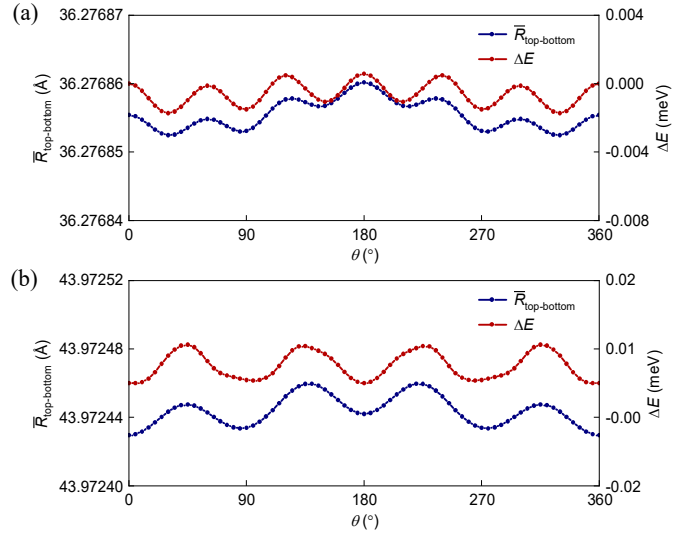

FIG. S9. Pairwise interlayer energy and average interatomic distance between atoms in the top BP disk and in the bottom BP disk, as a function of disorientation angle, for (a) BP disks with $D$ of 60 Å and separated by 20 Å, (b) BP disks with $D$ of 80 Å and separated by 20 Å.

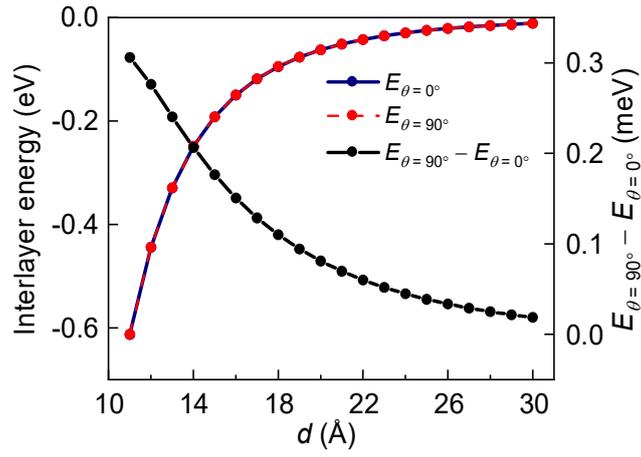

FIG. S10. Interlayer MBD energies on a fixed disorientation angles (0º or 90º) and maximum interlayer MBD energy difference ($E_{90°} - E_{0°}$), as a function of separation, for circular BP disks with $D$ of 60 Å.

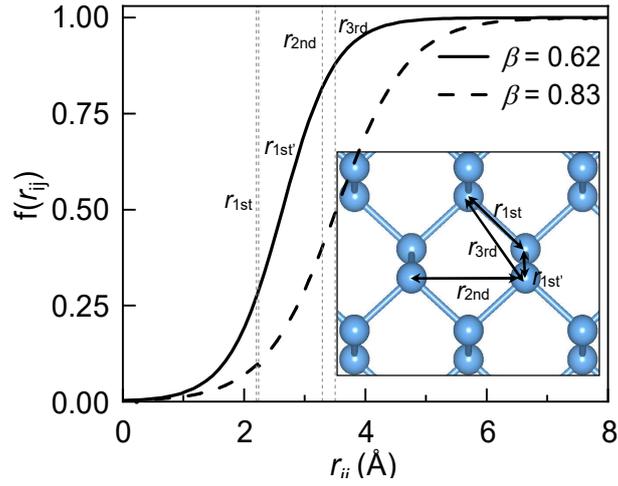

FIG. S11. Fermi-type damping functions with $\beta$ of 0.62 (solid line) and 0.83 (dashed line) used to truncate the short-range Coulomb potential. The inset shows the atomic structure of BP monolayer and the separations between the nearest atomic pairs. Self-consistent screening under the damping function with $\beta$ of 0.62 leads to an in-planar polarizability anisotropy in agreement with the HSE06 result. $\beta$ of 0.83 is the usual default setting in the MBD.

Table SI. Maximum MBD torque density between BP monolayers, in comparison with the Casimir-Lifshitz (CL) results for BP monolayers [12] and the experimental results for bulk materials [13].

| Materials | $M_{max}$ (nN·m·m$^{-2}$) | $d$ (nm) |
|:---:|:---:|:---:|
| BP-BP (MBD) | 487.38 | 2.00 |
| SiP$_2$-SiP$_2$ (MBD) | 1129.77 | 2.00 |
| Gr ribbon-BP (MBD) | 2082.37 | 2.00 |
| BP-BP (CL, $d^{-2.6}$) | 72.30 | 2.00 |
| BP-BP (CL, $d^{-4}$) | 1275.71 | 2.00 |
| 5CB-LiNbO$_3$ (exp) | 420 | 2.00 |
| 5CB-YVO$_4$ (exp) | 1077 | 2.00 |

Note: The maximum CL torque densities between BP monolayers at a separation of 2 nm are evaluated from Thiyam *et al.*'s CL result at a separation of 15.54 nm [12], according to the fitted power law of scaling ($M_{max} \sim d^{-2.6}$) shown in Fig. 2e of main manuscript or the ideal scaling law ($M_{max} \sim d^{-4}$) for strictly-2D layers. The experimental maximum torque density between 5CB and LiNbO$_3$ (or YVO$_4$) at a separation of 2 nm is evaluated from the experimental result at a separation of 18 nm [13], according to the ideal scaling law for semi-infinite plates ($M_{max} \sim d^{-2}$).

Table SII. In-planar polarizability anisotropy of BP and materials used in Somers *et al.*'s experiment [13].

| Materials | $(\varepsilon_{u0} - \varepsilon_{v0})/(\varepsilon_{v0} - 1)$ | $(\varepsilon_{u\infty} - \varepsilon_{v\infty})/(\varepsilon_{v\infty} - 1)$ |
|---|---|---|
| BP (HSE06) | - | 0.158 |
| SiP$_2$ (HSE06) | - | 0.277 |
| 5CB | 2.00[14] | 0.38[13] |
| LiNbO$_3$ | 0.53[15] | 0.15[13] |
| YVO$_4$ | 0.64[16] | 0.30[13] |

Table SIII. $C_6$ coefficient and polarizability coefficient of the materials, derived from the MBD-NL calculations.

| Atom | $C_6$ (hartree·bohr$^6$) | $\alpha$ (bohr$^3$) |
|---|---|---|
| P (BP) | 122.1 | 19.5 |
| Si (SiP$_2$) | 100.7 | 18.5 |
| P (SiP$_2$) | 129.5 | 20.8 |
| C (graphene) | 26.6 | 8.5 |